\documentclass[final]{svjour2}
\usepackage{graphicx}
\usepackage{rotating}
\usepackage{amssymb}
\usepackage{mathptmx}
\usepackage[numbers]{natbib}
\makeatletter
\journalname{Journal of Low Temperature Physics}
\begin{document}

\newcommand{\hdblarrow}{H\makebox[0.9ex][l]{$\downdownarrows$}-}
\title{On the Transition to Turbulence of Oscillatory Flow of Liquid Helium-4 }

\author{W. Schoepe}

\institute{Fakult\"at f\"ur Physik, Universit\"at Regensburg, D-93040
Regensburg, Germany\\
\email{wilfried.schoepe@physik.uni-regensburg.de}}

\date{14.12.2007}

\maketitle


\begin{abstract}

Oscillating solid bodies have frequently been used for studying the 
properties of normal and superfluid helium. In particular, the transition
from laminar flow to turbulence has attracted much interest recently. The purpose 
of this note is to review several central features of this transition in oscillatory flow, 
which have been inaccurately formulated in some recent work.

\end{abstract}

\section{Introduction}

Various types of oscillating bodies have been used for investigating 
the properties of liquid helium: vibrating wires, oscillating 
disks, grids, spheres and, recently, quartz tuning forks \cite{qfs06}. 

At low velocity amplitudes of the oscillating body the flow is laminar.
In this regime the amplitude grows linearly with the driving force. In the
superfluid the drag force is given by the interaction with the
quasiparticles and one can distinguish between a hydrodynamic and,
at low temperatures, a ballistic interaction, depending on the ratio of
the quasiparticle mean free path to the size of the body.
In case of a sphere, the hydrodynamic drag force is known quantitatively from Stokes' 
formula. For vibrating wires an approximation by a straight 
cylinder is made to allow for a quantitative analysis using Stokes' formula 
for a cylinder. For the more complicated geometry of a grid or a 
tuning fork the analysis is more qualititative. Nevertheless, the tuning fork 
and the vibrating wire have proven to be very useful secondary thermometers
because the damping force scales with the quasiparticle density. In the ballistic regime the drag is given simply by the scattering of quasiparticles off some geometrical cross section. Finally, between the hydrodynamic and the ballistic regime slip effects become important and a quantitative analysis is complicated. In the normal liquid the drag force in the laminar regime is classical and therefore of lesser interest here.

At large velocity amplitudes the flow becomes turbulent on either side of the oscillating body extending into a volume determined at least by the size and the amplitude of the body or even filling the entire measuring cell. 
In this regime the amplitude is found to grow with the square root of the driving force, both in the normal liquid and in the superfluid.
This signals a drag force that increases with the square of the velocity amplitude,
similar to classical fully developed turbulence for steady motion of the body. For a sphere the 
turbulent drag force in superfluid $^4$He was studied in detail \cite{prl, physica}.
Turbulent drag originates from the superfluid component. Only near the lambda temperature nonlinear drag of the normal component begins to contribute but
full turbulence in the normal phase was not observed. With the tuning forks, however, full turbulent
drag in the normal liquid was reached at sufficiently high velocities \cite{skrpre,skrjltp}.
The transition from laminar oscillating flow to turbulence is of great interest, not only in 
the superfluid regime, where shedding of quantized vortices will occur at some critical velocity \cite{tsub}, but 
also in the normal liquid because little is known experimentally about the different drag force when compared with
steady motion. In the following Section 2, elementary properties of the classical case will be summarized and compared with recent work. This will then be the basis of a discussion of results obtained in liquid $^4$He in Section 3.

\section{Transition to Turbulence of Classical Oscillatory Flow}

There are three independent length scales in oscillatory flow: the size $L$ of the oscillating body, 
the oscillation amplitude $A$, and the viscous penetration depth $\delta=\sqrt{2\nu/ \omega}$,
where $\nu$ is the kinematic viscosity of the liquid and $\omega$ is the oscillation frequency.
From these 3 quantities one can obtain 2 independent dimensionless numbers. Because the transition to turbulence is due to the nonlinear term of the Navier-Stokes equation when the velocity {\bf v} becomes large, the following 2 numbers are relevant \cite{landau}: 
\begin{enumerate}

\item The ratio of the nonlinear term $(\mathrm{{\bf v}}\nabla) \mathrm{{\bf v}} \sim \mathrm{{\bf v}}^2/ L$ to the viscous term $\nu\nabla^2 \mathrm{{\bf v}} \sim \nu \mathrm{{\bf v}} /L^2$. 
This ratio is the Reynolds number $Re=L\mathrm{{\bf v}}/\nu$. Inserting $\mathrm{{\bf v}}=\omega A$ gives $Re=2LA/\delta^2
$.

\item The ratio of the nonlinear term to the time derivative $\dot{\mathrm{\bf v}}\sim A \omega^2$. This is called the Strouhal number $Sr$ in Ref.\cite{landau} which is given here by $Sr=A/L$. Note that the Strouhal number does not depend on the viscosity and hence remains valid in an ideal liquid.

\end{enumerate} 
Thus, the situation is more complicated than in steady flow where only the Reynolds number is relevant. Now, one has to take into account the Strouhal number as well. There are 2 limiting cases, though, where the situation is simpler  \cite{landau}:

\begin{enumerate}
\item If $L\ll\delta$, i.e., at low frequencies or low temperatures in superfluid $^4$He (but still in the hydrodynamic regime), where $\delta$ of the normal component diverges, the time dependence of {\bf v} can be neglected. The situation is analogous to stationary flow and consequently the nonlinear term is negligible if $Re\ll1$.\\

\item If $L\gg\delta$, i.e., at large frequencies or in $^4$He close to or above $T_\lambda$, the nonlinear term can be neglected only if the $Sr\ll1$. The viscous term is much smaller than $A\omega^2$ and the Reynolds number $2AL/\delta^2$ need not to be small.

\end{enumerate}

It is clear that no pair of numbers other than $Re$ and $Sr$ is relevant for the transition to turbulence of oscillatory flow. For example, the choice to consider the Keulegan-Carpenter number ($Kc$) and the Stokes number ($St$) instead \cite{hann}, is not useful. While $Kc = 2 \pi A/L$ is essentially the same as the Strouhal number\footnote{In classical fluid dynamics often a different definition of the Strouhal number is used and the definition here is called Keulegan-Carpenter number.}, the Stokes number is $St=\omega L^2/\nu=2L^2/\delta^2$. Because $L/\delta$ may be either a large number or a small one in the laminar regime, no information on the transition to turbulence can be inferred from its value. The reason for this fact is that $St$ is given by the ratio of the Reynolds number $2AL/\delta^2$ to the Strouhal number $A/L$. Hence, it does not depend on the nonlinear term, which drops out. It just compares the 2 linear terms of the Navier-Stokes equation: it is given by the ratio of $\dot{\mathrm{\bf v}}\sim A \omega^2$ to the viscous term $\nu\nabla^2 \mathrm{\bf v} \sim \nu \mathrm{A \omega} /L^2$.

Moreover, it is also clear that one cannot replace $L$ by $\delta$ in $Re$ and $Sr$ as suggested recently when discussing experiments with tuning forks on the transition to turbulence in liquid helium \cite{skrpre,skrjltp}, see below. This would leave only $A$ and $\delta$ as independent length scales from which only one dimensionless number, namely $A/\delta$, can be obtained. This would give $Re=2A/\delta$ and $Sr=A/\delta$ which obviously does not make sense because by definition both numbers refer to independent ratios of the nonlinear term to the linear ones, and also $Sr$ would then depend on the viscosity.

\section{Transition to Turbulence in Liquid Helium}

Recent experiments with tuning forks on the transition from laminar flow to fully developed turbulence cover a temperature range from 1.3 K up to 4.2 K for the liquid and include also gaseous helium at 78 K \cite{skrpre,skrjltp}. In the normal liquid and in the gas a gradual transition from the linear behavior of {\bf v} as a function of the drive $F$ to the square root dependence is observed, extending over more than 2 orders of magnitude in {\bf v}. This is similar to what is known from steady motion, see, e.g., Ref. \cite{landau}. By extrapolating both regimes a "critical" velocity {\bf v}$_{cr}$ can be defined where the laminar drag force $F_{drag}^{lam}=\lambda${\bf v} becomes equal to the turbulent drag force $F_{drag}^{turb}=\gamma$\,{\bf v}$^2$, hence ${\bf v}_{cr}=\lambda/\gamma$. The drag coefficients $\lambda$ and $\gamma=C_d\rho\sigma/2$ can be obtained from the data ($C_d$ is a numerical drag factor that depends on the geometry of the body, $\rho$ is the density of the liquid, and $\sigma$ is the area normal to the flow).\footnote{To be precise, for nonlinear drag forces the principle of energy balance (energy gain from drive = energy loss from drag) has to be applied for inferring the drag force from the driving force. But this will introduce only a numerical factor of order one in $\gamma$ which is not important here \cite{physica}.} This "critical" velocity is found to scale as $\sqrt{\nu\omega}$ \cite{skrpre,skrjltp}. For a sphere the drag coefficient $\lambda$ is known analytically (Stokes' solution) and therefore the validity of this scaling can be proven \cite{skrpre}. For the geometry of a fork, $\lambda$ obviously has a similar dependence on $\nu$ and $\omega$ and differs only by a numerical factor. It is important to note, that this scaling behavior is understood without postulating, as was done in Refs. \cite{skrpre,skrjltp}, that the length $L$ in the Reynolds number should be replaced by the penetration depth $\delta$.  Finally, in view of the wide interval between the laminar regime and full turbulence, extending over more than 2 orders of magnitude, it appears questionable that this "critical" velocity has any physical meaning. At least for steady flow around a sphere or a cylinder it has none that I know of.

In the superfluid phase of $^4$He the situation is quite different. Well below $T_\lambda$ where the normal fluid density is small compared to the superfluid density, the transition to turbulence is sharp, even discontinuous or hysteretic \cite{prl,physica,yano1,pickett0}. The reason is that turbulence in the superfluid phase occurs abruptly at a critical velocity when vorticity is created. The classical Strouhal number is obviously not applicable for the transition to turbulence in a quantum fluid. 

Experiments with a sphere \cite{olli} show that the turbulent drag force is no longer proportional to {\bf v}$^2$ but instead has a constant shift $F_{drag}^{turb}=\gamma$\,{\bf v}$^2-F_0$. This implies that superfluid turbulent drag sets in at a temperature independent critical velocity ${\bf v}_{cr}=\sqrt{F_0/\gamma}$. Up to now, no theory exists that could describe the measured critical velocities quantitatively, probably because both geometry and surface properties of the oscillating body will have an influence. Furthermore, careful experiments with a vibrating wire show that the critical velocity may be affected by remanent vortices \cite{tsub,hashi}. The normal phase may remain in the laminar regime or gradually start deviating from it at high velocity amplitudes. With the sphere, e.g., very little nonlinear normal drag is found near $T_\lambda$ \cite{prl, physica}, which is reasonable because $Sr \le 0.5$ in this case. The implication is that quantum turbulence in the superfluid phase of $^4$He does not necessarily lead to classical turbulence in the normal phase, in contrast to what seems to be generally believed. 

Experiments with an oscillating grid \cite{Mc1} indicate that the resonance frequency is lower when turbulence is produced. This is attributed to an enhanced effective mass of the grid due to vortices attached to it. However, with wires the opposite behavior is found \cite{yano1,pickett0}. Very recently, a decrease of the resonance frequency of the wire is reported in a vortex free environment, and an increase if remanent vorticity is present \cite{hashi}. (In the latter case the spring constant of the oscillator is believed to be increased by vortices attached to the body and the walls of the cell.)  Skepticism is expressed with regard to the interpretation of the enhanced mass effect \cite{hann,Mc2}. With the sphere no change of the resonance frequency due to turbulence production was detectable within a resolution of better than $10^{-4}$. This situation clearly deserves further investigation. 

In the ballistic regime, at very low temperatures, the flow pattern switches intermittently from laminar flow (better: potential flow) to turbulence and back \cite{pickett0,olli,yano2}. The lifetimes of the turbulent phases are exponentially distributed and can be attributed to statistical fluctuations of the vorticity \cite{prl2}. Turbulence appears to be stable when sufficient power from the drive is available so that the turbulent lifetimes exceed the measuring time.\footnote{Because the mean turbulent lifetime is found to increase exponentially with the power input from the drive but never really diverges, superfluid turbulence is inherently unstable. It only appears to be stable because of the finite observation time of the turbulent phases (up to 100 minutes in Ref. \cite{olli}). Interestingly,  also in classical turbulence an inherent instability has recently been observed in pipe flow \cite{hof}.} The lifetimes of the laminar phases, however, which have a Rayleigh distribution \cite{olli}, remain to be explained. Finally, metastable potential flow is observed above ${\bf v}_{cr}$ whose lifetime is limited only by natural radioactivity, which may create local vorticity in the superfluid that triggers the decay of the potential flow \cite{olli}. Also this observation of metastable potential flow awaits a theoretical interpretation.

\section{Summary}
Several problems with the interpretation of experiments on the transition to turbulence of oscillatory flow of $^4$He have been discussed. Some have been resolved by reviewing the classical case. In the superfluid, some remain open for further investigation. Corresponding experiments in superfluid $^3$He remain beyond the scope of the considerations here, because the situation is quite different: pair breaking and Andreev scattering of quasiparticles by the superflow contribute to the drag on the oscillating body, and the strength of mutual friction determines whether or not turbulence can exist. 

\begin{acknowledgements}

It is a pleasure to thank Matti Krusius and his team at Helsinki University of Technology for warm hospitality, for helpful discussions, and for the priviledge to participate in their experiments with tuning forks in superfluid helium.

\end{acknowledgements}

\end{document}